\documentclass[aps,twocolumn,raggedbottom,superscriptaddress,tightenlines,prl,10pt]{revtex4-1}
\usepackage{amsthm}
\usepackage{amsmath}
\usepackage{amssymb}
\usepackage{multirow}
\usepackage{graphicx}
\usepackage{epstopdf}
\usepackage{url}
\usepackage{float}
\usepackage{bm}
\usepackage{ifthen}
\usepackage[usenames,dvipsnames]{color}
\usepackage{mathrsfs}
\usepackage{natbib}
\usepackage[colorlinks=true,citecolor=blue,urlcolor=blue]{hyperref}
\usepackage{epstopdf}
\usepackage{braket}
\usepackage{array}
\usepackage{siunitx}

\makeatletter
\newcommand{\thickhline}{%
    \noalign {\ifnum 0=`}\fi \hrule height 1pt
    \futurelet \reserved@a \@xhline
}
\newcolumntype{"}{@{\hskip\tabcolsep\vrule width 1pt\hskip\tabcolsep}}
\makeatother

\newcommand{\epscor}{\varepsilon_{\text{cor}}}
\newcommand{\epssec}{\varepsilon_{\text{sec}}}

\newcommand{\ebit}{e_{\text{bit}}}
\newcommand{\eph}{e_{\text{phase}}}

\graphicspath{{Images/}} % Put all images in this directory to avoid clutter
\begin{document}

\newcommand{\MITPhysics}{\affiliation{Department of Physics, Massachusetts Institute of Technology, Cambridge, Massachusetts 02139, USA}}
\newcommand{\RLE}{\affiliation{Research Laboratory of Electronics, Massachusetts Institute of Technology, Cambridge, Massachusetts 02139, USA}}
\newcommand{\Sandia}{\affiliation{Sandia National Laboratories, Albuquerque, New Mexico 87123, USA}}
\newcommand{\UNM}{\affiliation{Department of Electrical \& Computer Engineering, University of New Mexico,  Albuquerque, New Mexico 87131, USA}}
\newcommand{\LL}{\affiliation{MIT Lincoln Laboratory, Lexington, Massachusetts 02421, USA}}
\pagestyle{empty}
\title{Metropolitan quantum key distribution with silicon photonics}

\author{Darius Bunandar}
\email{dariusb@mit.edu}
\RLE \Sandia
\author{Anthony Lentine}
\Sandia
\author{Catherine Lee}
\RLE \LL
\author{Hong Cai}
\Sandia
\author{Christopher M. Long}
\Sandia
\author{Nicholas Boynton}
\Sandia
\author{Nicholas Martinez}
\Sandia
\author{Christopher DeRose}
\Sandia
\author{Changchen Chen}
\RLE
\author{Matthew Grein}
\LL
\author{Douglas Trotter}
\Sandia
\author{Andrew Starbuck}
\Sandia
\author{Andrew Pomerene}
\Sandia
\author{Scott Hamilton}
\LL
\author{Franco N. C. Wong}
\RLE
\author{Ryan Camacho}
\Sandia
\author{Paul Davids}
\Sandia
\author{Junji Urayama}
\Sandia
\author{Dirk Englund}
\RLE
\date{\today}

\begin{abstract}
Photonic integrated circuits (PICs) provide a compact and stable platform for quantum photonics. Here we demonstrate a silicon photonics quantum key distribution (QKD) transmitter in the first high-speed polarization-based QKD field tests. The systems reach composable secret key rates of 950 kbps in a local test (on a 103.6-m fiber with a total emulated loss of 9.2~dB) and 106 kbps in an intercity metropolitan test (on a 43-km fiber with 16.4~dB loss). Our results represent the highest secret key generation rate for polarization-based QKD experiments at a standard telecom wavelength and demonstrate PICs as a promising, scalable resource for future formation of metropolitan quantum-secure communications networks.
\end{abstract}
\maketitle

%%%%%%%%%%%%%%%%
% INTRODUCTION

Quantum key distribution (QKD) remains the only quantum-resistant method of sending secret information at a distance~\cite{Tittel2002a,Scarani2009a}. The first QKD system ever devised used polarization of photons to encode information~\cite{BB84,Bennett1992}. QKD has since progressed rapidly to several deployed systems that can reach point-to-point secret key generation rates in the upwards of 100~kbps~\cite{Sasaki2011,Yoshino2013a,Dixon2015,Lee2016} and to other photonic degrees of freedom: time~\cite{Comandar2015,Korzh2014c,Tang2014b,Sibson2017}, frequency~\cite{Lee2014,Ali-Khan2007,Zhong2015b,Nunn2013}, phase~\cite{Inoue2002}, quadrature~\cite{Zhang2008,Fossier2009a,Jouguet2013,Pirandola2015}, and orbital angular momentum~\cite{Mafu2013}. While polarization remains an attractive choice for free-space QKD due to its robustness against turbulence~\cite{Sergienko:97,Buttler1998,Waks2002,Hughes2002,Vallone2014,Bourgoin2015}, polarization is commonly thought to be unstable for fiber-based QKD. For this reason, there has been a strong interest in translating the polarization QKD components into photonic integrated circuits (PICs), which provide a compact and phase-stable platform capable of correcting for polarization drifts in the channel. Recently, silicon-based polarization QKD transmitters were used for laboratory QKD demonstrations~\cite{Ma2016,Sibson2016}, but their performance advantage over standard telecommunication components has yet to be demonstrated. Here we report the first field tests using high-speed silicon photonics-based transmitter for polarization-encoded QKD.

The silicon photonics platform allows for the integration of multiple high-speed photonic operations into a single compact circuit~\cite{Reed2005,Michel2010,Leuthold2010,Baehr-Jones2012a}. Operating at gigahertz bandwidth, a silicon photonics polarization QKD transmitter can correct for polarization drifts with typical millisecond time scales in a metropolitan-scale fiber link. Furthermore, silicon nanophotonic devices are compatible with the existing complementary metal-oxide-semiconductor (CMOS) processes that have enabled monolithic integration of photonics and electronics, possibly leading to future widespread utilization of QKD.

The QKD transmitter demonstrated here is manufactured using a CMOS-compatible process. The transmitter combines a 10-Gbps Mach-Zehnder Modulator (MZM) with interleaved grating couplers, which convert the polarization of a photon in an optical fiber into the path the photon takes in the integrated circuit, and vice versa. The high-speed polarization control is enabled by electro-optic carrier depletion modulation within the MZM~\cite{Cai2017}. We show the performance of the device in a local field test and an intercity field test. With a clock rate of 625~MHz, we generated secret keys at a rate of 950~kbps and observed a bit error rate of $2\%$ in the local test between two neighboring buildings connected by a 103.6~m fiber (with an additional 9~dB emulated loss). In the 43~km (16.4~dB channel loss) intercity test between the cities of Cambridge and Lexington, we generated secret keys at a rate of 106~kbps and observed a bit error rate $2.8\%$. Both QKD operations are demonstrated to be secure against collective attacks in a composable security framework with a tight security parameter of $\epssec = 10^{-10}$. Our results demonstrate how silicon photonics---supported by the currently existing CMOS technology---can pave the way for a high-speed metropolitan-scale quantum communication network.

%%%%%%%%%%%%%%%%
\begin{figure*}[t]
    % \centering
    \includegraphics[width=\textwidth]{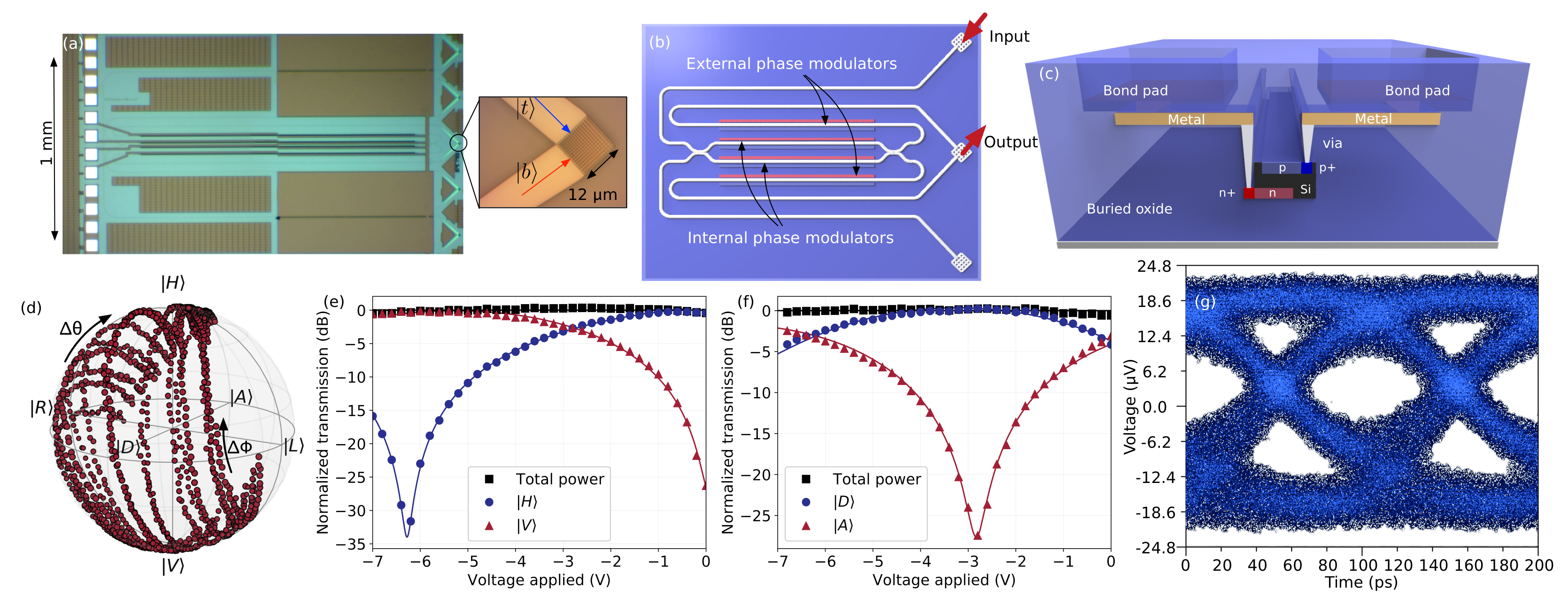}
    \caption{
    (a) Optical micrograph of the silicon photonics transmitter, along with a scanning electron micrograph of the polarization grating coupler. Only the inner three polarization grating couplers are parts of the transmitter operation; the outer two couplers are present to help alignment with a fiber v-groove array.
    (b) Schematic diagram of the MZM transmitter. The device uses two internal and two external electro-optic phase modulators, each of length 1.5~mm.
    (c) Schematic of the cross-sectional layer stack of the transmitter.
    (d) Bloch sphere representation of the polarization states generated by the transmitter as the internal ($\Delta \theta$) and the external ($\Delta \phi$) phase modulators are biased.
    (e--f) Polarization modulation with the silicon photonics polarization modulator as measured in the two relevant bases. Polarization extinction ratio of more than 25~dB can be typically achieved. Negative voltage denotes reverse bias with regards to the doped p-i-n junction. Measurements in the $Z$-basis and $X$-basis are shown in (e) and (f), respectively.
    (g) Eye diagram of 10~Gbps polarization modulation in the $Z$-basis: on-state corresponds to $\ket{V}$ and off-state corresponds to $\ket{H}$. 
    }
	\label{fig:device}
\end{figure*}
%%%%%%%%%%%%%%%%

\section*{Results} % (fold)
\label{sec:results}

\subsection*{Silicon photonics transmitter} % (fold)
\label{sub:silicon_photonics_transmitter}

%%%%%%%%%%%%%%%%
% DESCRIPTION OF TRANSMITTER, ALONG WITH QUBIT VERIFICATION

Our QKD transmitter and its cross-section are shown in Fig.~\ref{fig:device}(a--c). Light is coupled in and out of the transmitter using a standard fiber v-groove array of \SI{250}{\micro\meter} pitch. Owing to the large index contrast between the silicon layer and the buried oxide, the transmitter is compact within a total area of $0.75 \times 1.5$~\si{mm^2}. Polarization grating couplers are used to convert between polarization-encoding in the input/output fibers and path-encoding within the PIC. The unitary transformation is similar to that of a polarizing beam splitter (PBS). Within the PIC, the photons' paths---and their relative phases---are manipulated using an MZM with two internal and two external electro-optic phase modulators, which in turn manipulate the photon polarization in the output fiber.

The input polarization grating coupler separates light from the horizontal and vertical polarizations onto two different paths, both in the transverse-electric (TE) polarization: with its electric field oscillating parallel to the chip surface. Any light inadvertently converted into the transverse-magnetic (TM) polarization in these waveguides is greatly attenuated by the phase modulators which strongly support higher transmission in TE polarization over TM polarization.

The electro-optic phase modulators in the MZM are based on depletion-mode free-carrier dispersion from a doped p-i-n junction superimposed on the optical mode~\cite{reed2004silicon,moss2013semiconductor}. The overlap between the optical mode and the free carriers results in free carrier refraction~\cite{Soref1987}, which can be controlled with gigahertz RF signals to achieve high-speed phase modulation.

The polarization states generated by the transmitter have a purity of $1.000 \pm 0.005$, measured using a polarimeter. In Fig.~\ref{fig:device}(d), the relative phases of the internal phase shifters ($\Delta \theta$) as well as the relative phases of the external phase shifters ($\Delta \phi$) are swept with a reverse bias voltage between 0 and 8~V. For this voltage range, the polarization states lie on the surface of the Bloch sphere indicating that they remain pure throughout.

The BB84 QKD protocol requires Alice to prepare three quantum states: two eigenstates of $Z$ and an eigenstate of $X$~\cite{Tamaki2014,Mizutani2015}. Alice randomly chooses the basis she prepares in. When $Z$-basis is selected, Alice prepares either $\ket{0_z} = \ket{H}$ or $\ket{1_z} = \ket{V}$ with equal probabilities of 1/2. Otherwise, when $X$-basis is selected, Alice prepares the state $\ket{0_x} = \ket{D} = (\ket{H} + \ket{V})/\sqrt{2}$. 

We prepared the three quantum states at high fidelity, as shown in Figs.~\ref{fig:device}(e--f), with a polarization extinction ratio better than 25~dB which is required for low-error QKD operations. The internal and external phase modulators were configured to produce the state $(\ket{t}+\ket{b})/\sqrt{2}$, which we take to be $\ket{0_z}$. RF signals of differing voltages were applied to one of the external phase modulators to generate $(\ket{t}+e^{i\phi}\ket{b})/\sqrt{2}$, where $\phi$ is the applied phase shift. All the three BB84 states can be generated by applying the phase shifts $\phi = 0$, $\pi/2$, and $\pi$. The polarization states were measured using a PBS followed by two InGaAs photodiodes. A polarization controller before the PBS allowed measurements in the two BB84 bases: the $Z$-basis and the $X$-basis.

As shown in the eye diagram in Fig.~\ref{fig:device}(g), the phase modulators allowed us to generate the polarization states at 10~Gbps. These measurements were acquired by using an in-line polarizer placed at the output of the transmitter that converts the polarization state $\ket{V}$ into an on-state and the polarization state $\ket{H}$ into an off-state. When the transmitter was modulated at 6~Gbps or lower, not a single error was observed for a five-minute operation. At a 10~Gbps data rate, as shown here, we measured a low error rate of $9.0 \times 10^{-10}$~\si{s^{-1}}.

% subsection silicon_photonics_transmitter (end)

\subsection*{Field tests} % (fold)
\label{sub:field_tests}

%%%%%%%%%%%%%%%%
% DESCRIPTION OF FIELD TEST SETUPS

%%%%%%%%%%%%%%%%
\begin{figure*}[t]
    % \centering
    \includegraphics[width=\textwidth]{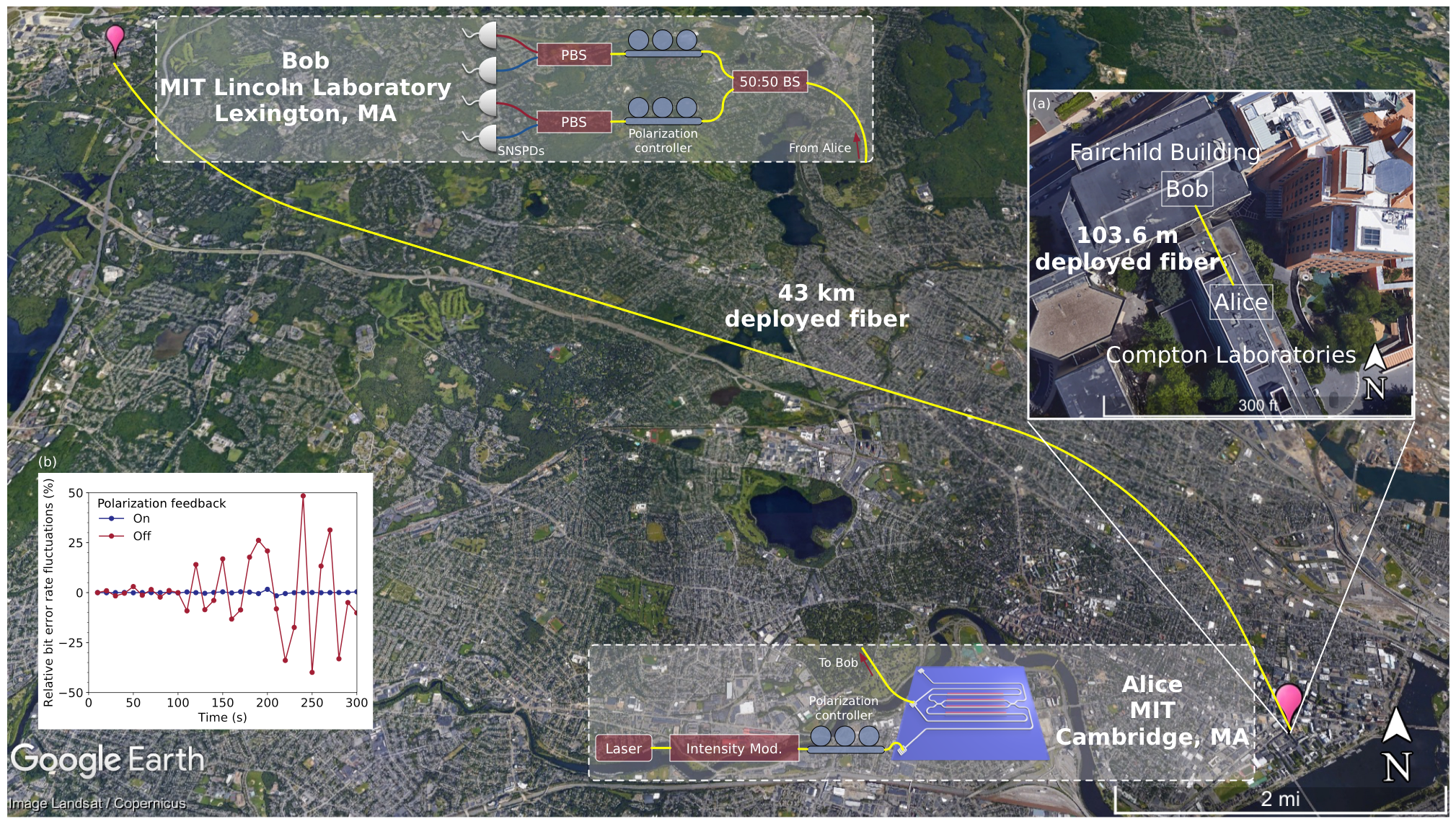}
    \caption{Aerial view of the intercity QKD field test. Alice is located at Massachusetts Institute of Technology (MIT) in Cambridge and Bob is located at MIT Lincoln Laboratory in Lexington. Although the point-to-point distance between the two stations are $\sim18$~km, they are connected by a 43-km dark fiber link. Alice consists of an attenuated laser source, an intensity modulator, and the silicon photonics polarization transmitter. Bob consists of two polarizing beam splitters (PBSs) followed by four superconducting nanowire single photon detectors (SNSPDs). \emph{Insets:} (a) Close-up aerial view of the local QKD field test, where Alice and Bob are located in two adjacent MIT buildings connected by a 103.6~m deployed dark fiber link. Alice and Bob's setups are the same as the ones used in the intercity test.  (b) Fluctuations on the bit error rate with and without polarization feedback control, relative to the starting bit error rate. Imagery \textcopyright2017 Google. Map data: Google, Landsat/Copernicus.}
	\label{fig:setup}
\end{figure*}
%%%%%%%%%%%%%%%%

We performed two QKD field tests: a local test and an intercity test. Fig.~\ref{fig:setup} shows a map of the greater Boston area, identifying the locations of Alice and Bob, together with the experimental setups implementing the asymmetric polarization-based BB84 protocol. Alice, located in the Compton Laboratories at MIT for both field tests, prepares the three polarization BB84 states at random. Bob measures in either the $Z$-basis or the $X$-basis using four superconducting nanowire single-photon detectors (SNSPDs) at a different location for each field test. He is located in the Fairchild Building for the local test and at MIT Lincoln Laboratory in Lexington for the intercity test. Bob makes his basis choices using the polarization controller placed before each PBS.

Alice creates (non-phase randomized) attenuated laser pulses of width 800~ps at 1480~nm with a 625~MHz repetition rate. The pulses are modulated into the three BB84 polarization states by the silicon photonics trasmitter. Alice first calibrates for the polarization rotation through the channel, and DC reverse voltage biases are applied to the transmitter such that the state $\ket{D}$ is generated by default. To generate the states $\ket{H}$ and $\ket{V}$, Alice applies synchronized RF pulses with a full-width-at-half-maximum of 400~ps. To maximize the length of secret keys generated, Alice chooses to prepare either in the $Z$-basis with a probability of 15/16 (and in the $X$-basis with a probability of 1/16).

In the local test, Alice sends her prepared states to Bob through a 103.6~m fiber link connecting the two laboratories. The loss through the link is $0.2$~dB, and we emulated longer fiber distances by installing a variable optical attenuator before the channel. On the other hand, the deployed intercity fiber connecting Cambridge and Lexington is 43~km long with 16.4~dB loss. 

Bob detects the pulses he receives in either of the two bases with $50$\% probability to maximize the number of security check events when the key-generating detectors are saturated. For the local test, Bob uses four individual WSi SNSPDs, each with a quantum efficiency greater than 85\%, a timing resolution of $\sim250$~ps, and a background dark count rate of $\sim1000$~counts/s. For the intercity test, Bob uses four NbN SNSPD systems, each consisting of two interleaved NbN nanowires with a single optical fiber input. A single NbN SNSPD system has a quantum efficiency of $30$\%, a timing resolution of tens of ps, and a background dark count rate of $\sim1000$~counts/s. While the WSi nanowire has a better quantum efficiency than the NbN system, the WSi detector saturates at a lower count rate of $5\times10^6$~counts/s.

Alice and Bob only generate secret keys when both parties choose the $Z$-basis. The quantum bit error rate $\ebit$ is measured by checking the number of bits that have been flipped between their raw bit strings. On the other hand, the quantum phase error rate $\eph$ can be estimated from $X$-basis events along with the mismatched basis events, where Alice and Bob chose different preparation and measurement bases (see Methods)~\cite{Tamaki2014,Mizutani2015}.

An automated polarization feedback system is placed in the intercity channel between Alice and Bob, which can drift significantly on the timescale of the experiment. To correct for the drift, Alice sends a series of calibration signals and optimizes her DC voltage biases such that the error rate on both measurement bases is kept low. As seen in Fig.~\ref{fig:setup}(b), the relative fluctuations of $\ebit$ (relative to its starting value) are limited to $2\%$ with feedback, and to about $50\%$ without feedback.

% subsection field_tests (end)

\subsection*{Composable secret key generation} % (fold)
\label{sub:composable_secret_key_generation}

\begin{figure}[t]
    % \centering
    \includegraphics[width=\columnwidth]{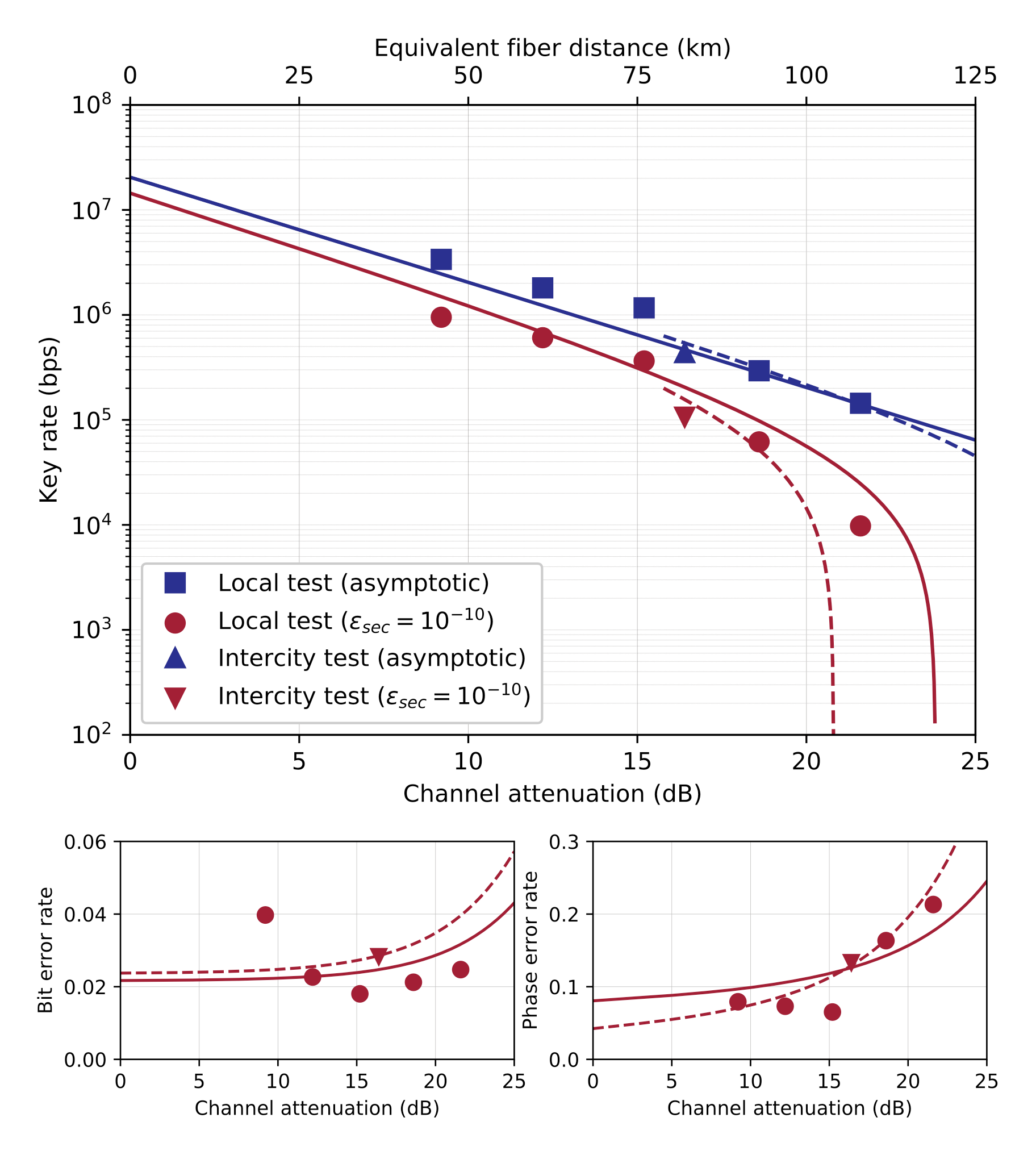}
    \caption{
    \emph{Top:} Experimental SKRs at different channel losses. The (blue) squares and upright triangles are asymptotic SKRs for the local test and the 43~km metropolitan intercity test, respectively. Similarly, the (red) circles and inverted triangles are the SKRs calculated within the composable security framework with $\epssec = 10^{-10}$ for the local test and the metropolitan field test, respectively. In the local tests, a variable attenuator is used to provide higher attenuation beyond the channel's 0.2~dB loss. Solid and dashed lines correspond to numerical simulations of the SKRs for the local test and the intercity test, respectively.
    \emph{Bottom:} Bit and phase error rates against channel loss in the composable security framework. The symbols used here are the same as the ones above.}
    \label{fig:keyRate}
\end{figure}

Figure~\ref{fig:keyRate} shows the performance of the QKD transmitter in both field tests, in terms of the observed secret key rate (SKR), $\ebit$, and $\eph$. For clarity, we plotted the SKRs against the channel loss and the equivalent fiber distance assuming an optimistic fiber loss of 0.2~dB/km. We kept the number of pulses sent from Alice to Bob at $N = 2.81\times10^{11}$ to maintain a uniform collection time of 450~s for each experiment, and analyzed the security with a small security parameter of $\epssec = 10^{-10}$.

For the local test, at a total channel attenuation of 9.2~dB, we obtained a SKR of 950~kbps using mean photon numbers of 0.12, 0.012, 0.003 for the signal and the two decoy states---chosen with probabilities 2/3, 2/9, and 1/9---respectively. The mean photon numbers were kept low to avoid detector saturation. The total channel attenuation was further increased from 9.2~dB to 24.2~dB to simulate longer fiber distances. We observed an average $\ebit$ of $\sim2\%$, except for the lowest channel attenuation where $\ebit$ is higher at $3.97\%$ as the WSi detectors are saturated. As expected from theoretical simulations, the phase error rate $\eph$ increased from $7.92\%$ to $21.31\%$ as we increased the channel attenuation.

For the metropolitan intercity test, we obtained an SKR of 106~kbps using mean photon numbers 0.5, 0.03, 0.015 for the signal and the decoy states with the same probabilities as above. Here the mean photon numbers could be chosen higher while being well under the NbN detector systems' saturation point. We observed an $\ebit$ of $2.82\%$ and an $\eph$ of $13.29\%$ in this 43~km experiment.

%%%%%%%%%%%%%%%%
% EXPERIMENTAL RESULTS

%%%%%%%%%%%%%%%%

% subsection composable_secret_key_generation (end)

% section results (end)

%%%%%%%%%%%%%%%%
% TABLE COMPARING OUR EXPERIMENT WITH OTHERS

\begin{table*}[t]
	\centering
	\begin{tabular}{|c|c|c|c|c|c|c|c|c|}
	\hline
	\multirow{2}{*}{Reference} & Clock rate & $\lambda$ & Fiber distance & Channel loss & Secret key rate & \multirow{2}{*}{Finite-key $\epssec$} & \multirow{2}{*}{Protocol} & \multirow{2}{*}{Notes} \\
	 & (MHz) & (nm) & (km) & (dB) & (kbps) & & &\\
    \thickhline
	\cite{Gordon1999} & 1000 & 850 & 4.2 & 9.24 & 130& Assumes asymptotic & B92 & Polarization, VCSELs \\
    \hline
	\cite{Tang2006} & 625 & 850 & 1 & 2.2 & 2100 & Assumes asymptotic & B92 & Polarization, VCSELs \\
    \hline
	\cite{Ma2016} & 10 & 1550 & -- & 0.0 & 0.95 & Assumes asymptotic & BB84 & Polarization, Si PIC\\
    \hline
	\cite{Sibson2016} & 1000 & 1550 & 20 & 4.0 & 329 & Assumes asymptotic & BB84 & Polarization, Si PIC\\
    \hline
    \cite{Sasaki2011} & 1000 & 1550 & 50$^{\dagger}$ & 14.5 & 304 & Assumes asymptotic & BB84 & Time-bin, long-term\\
    \hline
    \multirow{2}{*}{\cite{Yoshino2013a}} & \multirow{2}{*}{1000} & 1547.72 &\multirow{2}{*}{22$^{\dagger}$} & \multirow{2}{*}{12.6} & \multirow{2}{*}{230} & \multirow{2}{*}{Assumes asymptotic} & \multirow{2}{*}{BB84} & \multirow{2}{*}{Time-bin, long-term} \\
     & & 1550.92 & & & & & & \\
     \hline
    \cite{Dixon2015} & 1000 & 1550 & 45$^{\dagger}$ & 14.5 & 300 & $10^{-10}$ & BB84 & Time-bin, long-term \\
    \hline
    % \cite{Lee2016} & & 1559 & 43$^{\dagger}$ & 12.7 & 1200 & $10^{-10}$ & HD & Time-bin with 4~bins \\
    % \hline
	\multirow{2}{*}{This work} & \multirow{2}{*}{625} & \multirow{2}{*}{1480} & 0.1$^{\dagger}$ + var. att. & 9.2 & 950 & \multirow{2}{*}{$10^{-10}$} & \multirow{2}{*}{BB84} & \multirow{2}{*}{Polarization, Si PIC}\\
		& & & 43$^{\dagger}$ & 16.4 & 106 & & & \\
	\hline
	\end{tabular}	
	\caption{Comparison of high-rate polarization-based QKD experiments and other high-rate discrete-variable QKD field tests. Dagger $(^{\dagger})$ represents a deployed fiber link. VCSELS: vertical-cavity surface-emitting lasers.}
	\label{tab:keyRate}
\end{table*}

%%%%%%%%%%%%%%%%

\section*{Discussion} % (fold)
\label{sec:discussion}

To illustrate the progress entailed by our results, we summarize our work in Table~\ref{tab:keyRate}  along with recent demonstrations of high-speed polarization-based QKD. Our work represents the highest observed SKR for any polarization-based QKD operations at comparable channel losses, and it performs comparably to other state-of-the-art QKD field demonstrations. It is also the first demonstration of the asymmetric loss-tolerant BB84 QKD protocol with guaranteed security against collective attacks~\cite{Mizutani2015}. The silicon photonics platform has enabled us to design a compact transmitter with high-speed and high-fidelity operations using a CMOS-compatible process. This points to the possibility of low-cost and resilient QKD transmitters for metro-scale quantum-secure networks.

PICs offer opportunities for further integration for both the transmitter and the receiver and for closing possible security flaws and side-channel attacks. Dense wavelength-division multiplexing has been one major thrust in classical communications, and a compact solution is available in silicon photonics by using an array of add-drop ring resonators~\cite{Ding2014,Liu2014}. This scheme can be integrated with our current QKD transmitter design with only minimal changes in the footprint. Furthermore, single photon detectors have been integrated into silicon photonics~\cite{Najafi2014}, showing the possibility of a compact QKD receiver.

Moreover, the configurability of the silicon photonics platform allows for complex monitoring circuits that protect against side-channel attacks~\cite{Lo2014}. For example, a Trojan horse attack can be thwarted by placing watchdog detectors in our transmitter~\cite{Muller1997,Stucki2002}. Possible detector vulnerabilities, such as the detector blinding attack~\cite{Jain2011,Lydersen2010a}, can be eliminated using the measurement-device-independent configuration~\cite{Xu2014,Comandar2015,Lo2012}.

In conclusion, we have demonstrated short-range and metro-scale QKD field tests based on a silicon photonics transmitter, reaching secret key rates of 950~kbps and 106~kbps, respectively. These are the first field tests using PIC-based QKD transmitters. The PIC platform provides a compact and phase-stable platform for high-speed QKD that is well suited for further scaling by wavelength division multiplexing.

% section discussion (end)

\bibliography{References/QKD,References/reviews,References/pic,References/references}

% \appendix

%%%%%%%%%%%%%%%%
% ACKNOWLEDGMENTS

\section*{Acknowledgements} % (fold)
\label{sec:acknowledgements}

We thank Evan Gabhart (MIT), Zheshen Zhang (MIT), Nicholas Harris (MIT), and P. Ben Dixon (MIT Lincoln Laboratory) for their helpful suggestions and discussion. D.E. and D.B. acknowledge support from the Air Force Office of Scientific Research, the Office of Naval Research, and the Samsung Advanced Institute of Technology.

This work was partly supported by the U.S. Department of Energy through the Sandia Enabled Communications and Authentication Network using Quantum Key Distribution (SECANT QKD) Grand Challenge, and performed, in part, at the Center for Integrated Nanotechnologies, an Office of Science User Facility operated for the U.S. Department of Energy Office of Science.

Sandia National Laboratories is a multimission laboratory managed and operated by National Technology and Engineering Solutions of Sandia LLC, a wholly owned subsidiary of Honeywell International Inc. for the U.S. Department of Energy’s National Nuclear Security Administration under contract DE-NA0003525.

DISTRIBUTION STATEMENT A: Approved for public release: distribution unlimited.

This material is based upon work supported by the Office of the Assistant Secretary of Defense for Research and Engineering under Air Force Contract No. FA8721-05-C-0002 and/or FA8702-15-D-0001. Any opinions, findings, conclusions or recommendations expressed in this material are those of the author(s) and do not necessarily reflect the views of the Assistant Secretary of Defense for Research and Engineering.

% section acknowledgements (end)

\section*{Author Contributions} % (fold)
\label{sec:author_contributions}

D.B., R.C., J.U., and D.E. conceived and designed the experiments. A.L., C.D., and P.D. designed the silicon photonics transmitter. D.T., A.S., and A.P. fabricated the transmitter. H.C., C.M.L., N.B., and N.M. characterized the transmitter. C.C., M.G., S.H., and F.W. prepared the single photon detectors. D.B., C.L., and H.C. performed the experiments and analyzed the data. D.B. and D.E. prepared the manuscript.

% section author_contributions (end)

\section{Methods} % (fold)
\label{sec:methods}

\subsection*{Protocol Description} % (fold)
\label{sec:protocol_description}
We consider an asymmetric three-state BB84 protocol. In particular, Alice randomly selects to prepare a qubit in either the $Z$-basis or the $X$-basis with probabilities $p_Z^A$ and $p_X^A = 1-p_Z^A$, respectively. Similarly, Bob independently and randomly chooses to measure in either of the two bases with probabilities $p_Z^B$ and $p_X^B = 1-p_Z^B$. In our experiments, $p_Z^A = 15/16$, $p_X^A = 1/16$, $p_Z^B = 1/2$, and $p_X^B = 1/2$. The mean photon number of each laser pulse in the experiment is chosen randomly from three different settings: $\mu_1, \mu_2, \mu_3$. They satisfy the relation $\mu_1 > \mu_2 + \mu_3$ and $\mu_2 > \mu_3 \geq 0$.

\begin{enumerate}
	\item \emph{Preparation}: For each laser pulse, Alice randomly chooses the mean photon number $\braket{N} \in \{\mu_1, \mu_2, \mu_3 \}$ with probabilities $p_{\mu_1}, p_{\mu_2},$ and $p_{\mu_3} = 1-p_{\mu_1}-p_{\mu_2}$, respectively. Alice then selects the basis $a \in \{Z,X\}$ with probabilities $p_Z^A$ and $p_X^A = 1-p_Z^A$, respectively. If she has selected the $Z$-basis, then she randomly sends either $\ket{0_z} = \ket{H}$ or $\ket{1_z} = \ket{V}$ to Bob with equal probabilities. If the $X$-basis was selected, she sends the $\ket{0_x} = \ket{D}$ to Bob. She records the bit value of the state she has sent in $x$.

	\item \emph{Measurement}: Bob measures the signals he received in the measurement basis $b \in \{Z,X\}$ with probabilities $p_Z^B$ and $p_X^B = 1 - p_Z^B$, respectively. Bob performs the measurements with four single-photon detectors (one per basis). He then records his measurement as one of the four possible outcomes: $\{0,1,\emptyset,\bot\}$. 0 and 1 are the bit values ($H$ and $V$ in the $Z$-basis, and $D$ and $A$ in the $X$-basis), $\emptyset$ represents no detection, and $\bot$ represents a double detection. Bob records the outcome in $y$, and he assigns a random bit value if a double detection is observed.

	\item \emph{Basis reconciliation and sifting}: Alice and Bob announce their bases and intensity choices over an authenticated public channel. They then place their records into one of the following sets:
		\begin{itemize}
			\item Key-generation sets: \\$\mathcal{Z}_{\mu} = \left\{ i| a_i = b_i = Z, \braket{N_i} = \mu, y_i \neq \emptyset \right\}$,
			\item Security-check sets: \\
			$\mathcal{X}_{\mu} = \left\{i | a_i = b_i = X, \braket{N_i} = \mu, y_i \neq \emptyset \right\}$,
			\item Mismatched-basis sets: \\
			$\mathcal{Z}^{j}\mathcal{X}^{k}_{\mu} = \left\{i|a_i=Z, b_i= X, \braket{N_i} = \mu,\right.$ \\ $\left.  x_i = j, y_i = k \right\}$.
		\end{itemize}
	Steps 1--3 are repeated until the size of each set has reached a certain length previously agreed by both parties. Alice and Bob generate a raw key pair $(\mathbf{Z}_A, \mathbf{Z}_B)$ by choosing a random sample from the set $\mathcal{Z} = \cup_{\mu} \mathcal{Z}_{\mu}$. Following~\cite{Lim2014}, we generate secret keys from all intensity settings.

	\item \emph{Parameter estimation}: Alice and Bob then compute the bounds to the number of vacuum and single-photon events within the set $\mathcal{Z}$ using the security-check sets and the mismatched-basis sets. Next, they estimate the number of phase errors within the single-photon events, and check if the phase error rate $\eph$ is less than the predetermined threshold value $e_{\text{phase, tol}}$. If $\eph > e_{\text{phase, tol}}$, then they abort the protocol, otherwise they proceed.

	\item \emph{Postprocessing}: Alice and Bob perform error correction for $(\mathbf{Z}_A, \mathbf{Z}_B)$ over their authenticated public channel, revealing $\lambda_{\text{EC}}$ bits. To verify that they have identical secret keys, they compute a two-universal hash function that publishes $\lceil \log_2 1/\epscor \rceil$ bits. If the protocol passes all the above steps, they then perform privacy amplification to extract a secret key pair $(\mathbf{K}_A, \mathbf{K}_B)$ with each key of length $\ell$ bits.
\end{enumerate}
% section protocol_description (end)

\subsection*{Security analysis} % (fold)
\label{sec:security_analysis}

We consider the loss-tolerant asymmetric BB84 protocol in the composable security framework~\cite{Tamaki2014,Mizutani2015}. A QKD protocol is considered to be secure if it is both correct and secret. The protocol is secret when the pair of keys $\mathbf{K}_A$ and $\mathbf{K}_B$ are identical except for some small probability $\epscor$, i.e. $\Pr \left[ \mathbf{K}_A \neq \mathbf{K}_B \right] = \epscor$. The probability $\epscor$ is determined by the failure probability of the two-universal hash function. Furthermore, the protocol is secret if the quantum state $\rho_{K_A E}$ that describes the correlation between Alice's key and Eve's quantum system is $\epssec$-close to $\omega_{K_A} \otimes \rho_E$, where $\omega_{K_A}$ describes a uniform distribution of all bit strings. In other words,
\begin{equation}
	\frac{1}{2} \left\| \rho_{K_A E} - \omega_{K_A} \otimes \rho_E \right\| \leq \epssec.
\end{equation}

Within this composable security framework, the secret key length is
\begin{equation}
\begin{aligned}
	\ell \geq & \left\lfloor m_0^L + m_1^L \left[ 1-h(\eph^U)-\xi h(\ebit) \right] \right.\\
	& \left. - \log_2 \frac{4}{\epssec^2} - \log_2 \frac{2}{\epscor} \right\rfloor,
\end{aligned}
\end{equation}
where $h$ is the binary entropy function, $m_0^L$ and $m_1^L$ are the lower bounds to the number detections due to vacuum and single photons, respectively. $\eph^U$ is an upper bound to the phase error rate, which can be computed using the methods outlined in Ref.~\cite{Mizutani2015}. $\ebit$ is the quantum bit error rate for the key-generating basis, and $\xi$ represents the error correction inefficiency---set at 1.15 for our calculations. For simplicity, we set all 17 failure probabilities related to estimating $m_0^L$, $m_1^L$, and the number of phase errors $N_{\phi}$ as $\varepsilon = \epssec^2/17$.

% section security_analysis (end)
% section methods (end)

\end{document}